\documentclass[12pt,draftcls,onecolumn]{IEEEtran}
\usepackage{amsmath,amsfonts,amssymb,amsbsy,bm,paralist,theorem,color}
\usepackage{graphicx,algorithmic,algorithm,relsize}
\usepackage{multicol}
\usepackage{multirow}
\usepackage{subfigure}
\usepackage{stfloats}
\usepackage{setspace}
\usepackage{cite}
\newtheorem{lem}{Lemma}

\newcommand{\Rset}{\mathbb{R}}

\newcommand{\Cset}{\ensuremath{\mathbb{C}}}
\newcommand{\rank}{\rm{rank}}

\newcommand{\SINR}{{\rm{SINR}}}
\newcommand{\tr}{{\rm{Tr}}}
\newcommand{\Rank}{{\rm{rank}}}
\DeclareMathOperator*{\st}{s.t.}

\graphicspath{{fig/}}
\allowdisplaybreaks[3]

\definecolor{orange}{RGB}{255,107,0}
\definecolor{green}{RGB}{0,160,20}

\begin{document}
	\title{Intelligent User Clustering and Robust Beamforming Design for UAV-NOMA Downlink}
	\date\today
	\author{Yanqing Xu, \IEEEmembership{Member, IEEE}, Fang Fang, \IEEEmembership{Member, IEEE}, Donghong Cai, \IEEEmembership{Member, IEEE}, and Yi Yuan, \IEEEmembership{Member, IEEE}
		\thanks{
			Y. Xu is with the Department of Standard Patent and Pre-Research, Huawei Technologies Co. Ltd, Beijing, China. (e-mail: xuyanqing2@huawei.com)
			
			F. Fang is with the School of Electrical and Electronic Engineering, The University of Manchester, U.K.. (e-mail: fang.fang@manchester.ac.uk).
		
			D. Cai is with the College of Cyber Security, Jinan University, Guangzhou, 510632, China. (e-mail: donghong\_cai@foxmail.com).
	
			Y. Yuan is with the Wolfson School of Mechanical, Electrical and Manufacturing Engineering, Loughborough University, Loughborough LE11 3TU, U.K. (e-mail: y.yuan@lboro.ac.uk)}
}
	\maketitle

\begin{spacing}{1}
		\begin{abstract}
		In this work, we consider a downlink non-orthogonal multiple access (NOMA) network with multiple single-antenna users and multi-antenna unmanned aerial vehicles (UAVs). In particular, the users are spatially located in several clusters by following the Poisson Cluster Process and each user cluster is served by a hovering UAV with NOMA. For practical considerations, we assume that only imperfect channel state information (CSI) of each user is available at the UAVs. Based on this model, the problem of joint user clustering and robust beamforming design is formulated to minimize the sum transmission power, and meanwhile, guarantee the quality of service requirements of users. Due to the integer variables of user clustering, coupling effects of beamformers, and infinitely many constraints caused by the imperfect CSI, the formulated problem is challenging to solve. For computational complexity reduction, the original problem is divided into user clustering subproblem and robust beamforming design subproblem. By utilizing the users' position information, we propose a k-means++ based unsupervised clustering algorithm to first deal with the user clustering problem. Then, we focus on the robust beamforming design problem. To attain insights on solving the robust beamforming design problem, we firstly investigate the problem with perfect CSI, and the associated problem is shown can be solved optimally. Secondly, for the problem in the general case with imperfect CSI, a semidefinite relaxation (SDR) based method is proposed to produce a suboptimal solution efficiently. Moreover, we provide a sufficient condition under which the SDR based approach can guarantee to obtain an optimal rank-one solution, which is theoretically analyzed. Finally, an alternating direction method of multipliers based algorithm is proposed to allow the UAVs to perform robust beamforming design in a decentralized fashion efficiently. Simulation results demonstrate the efficacy of the proposed algorithms and transmission scheme.
	\end{abstract}
\end{spacing}

\newpage


\section{Introduction}
To support the exponentially growing data traffic and number of devices in future wireless network, non-orthogonal multiple access (NOMA) was proposed to serve multiple users simultaneously on a same resource block \cite{forecast-2017,noma-survey,ding-jsac-2017}. 
The key idea of NOMA is to combine the superposition coding at transmitter and successive interference cancellation (SIC) at receiver, as such the spectral efficiency of wireless systems can be significantly improved \cite{tse-2005,XU-TSP-2017}. Due to its capability to provide superior spectral efficiency and massive connectivity, NOMA has been applied to many aspects of wireless communication systems, e.g., the Internet of Things network \cite{wang-twc-iot}, and the ultra-reliable and low-latency communications network \cite{xu-twc-urllc}. 
User clustering, also named as user grouping/pairing, is one fundamental issue of NOMA.
The impacts of user clustering on the system performance have been intensively investigated from both performance analysis and system design perspectives.
For example, \cite{ding-user-paring} studied the influences of user pairing on the outage probabilities of users in the NOMA system. \cite{cui-user-pairing} proposed to use a branch-and-bound based algorithm to solve the joint user pairing and power allocation optimally with the worst case computation complexity of NP-hard. To develop a computation-efficient algorithm for the user clustering problem, matching theory based heuristic algorithms were presented in \cite{cui-user-pairing,di-user-pairing,xu-tcom-harq}. Recently, to further reduce the computational complexity, \cite{cui-user-pairing} proposed to use the k-means based approach to perform unsupervised user clustering by exploring the users' position information.

Recently, unmanned aerial vehicles (UAVs) assisted wireless communications has received
considerable attentions due to its advantages to provide real time and high throughput services \cite{zeng-uav}. 
Compared to the conventional terrestrial wireless communication systems, the development of UAV
has also created a fundamental paradigm shift to facilitate fast and highly flexible deployment of
communication infrastructures. Specifically, by exploring the high maneuverability of UAV,
communication links can be established ubiquitously, especially in temporary hotspots, disaster
areas, and complex terrains \cite{sun-uav-mag}. Due to the above benefits, numerous efforts have been endeavoured to the research of UAV assisted wireless communication design. For example, \cite{wu-uav} proposed a suboptimal algorithm to solve the joint user association, UAV's trajectory and power allocation design. \cite{sun-uav} considered a solar-powered UAV system and a monotonic optimization based algorithm was developed to find the optimal UAV's trajectory and power allocation.
To benefit the advantages of multi-antenna technique, \cite{jiang-uav} investigated the suboptimal beamforming design and UAV positioning to maximize the throughput of UAV assisted network. Considering the advantages of NOMA and UAV, the application of NOMA to the UAV assisted network was recently investigated in \cite{tang-uav,nasir-uav}. However, all the aforementioned works assumed that the channel
state information (CSI) of the system can be perfectly acquired by the UAVs. In practice, due to,
e.g., imperfect channel estimation and finite feedback, the UAVs can never have perfect CSI.
Moreover, the wind would incur non-negligible body jittering of the UAVs, which will also harm
the acquisition of CSI \cite{choi-uav}. Therefore, it is of great importance to investigate the robust system design of UAV assisted wireless communication system under imperfect CSI assumptions. The robust beamforming design has been well studied in the orthogonal multiple access systems, e.g., \cite{shen-tsp,wang-twc,cai-twc}, and recently, has been extended to the NOMA system, e.g., \cite{su-tcom,sun-tcom,yuan2019}. More recently, the robust beamforming design has also been introduced to the UAV enabled system, e.g., \cite{pan-uav,new-uav}. Specifically, in \cite{pan-uav}, physical layer networking coding and multiple user detection approaches were proposed to combat the effects due to the absence of CSI, and \cite{new-uav} analyzed the power allocation problem in the single-antenna UAV-NOMA systems. 

In this paper, we study the joint user clustering and robust beamforming design problem of a downlink multi-user multi-UAV NOMA network with imperfect CSI assumption, where the users are spatially located by flowing the Poisson Cluster Process (PCP). Due to the integer variables relevant to user clustering, coupling effects of the downlink beamformers, and the infinitely many constraints bringing by the imperfect CSI, the resultant problem is challenging to solve. For computational complexity reduction, the original problem is divided into the user clustering subproblem and robust beamforming design subproblem. For the user clustering subproblem, unlike \cite{di-user-pairing,cui-user-pairing,liang2017,xu-tcom-harq}, where traditional optimization approaches are used, we would like to resort to unsupervised learning based approaches to solve the user clustering problem by exploring the users' position information. In particular, the users locates closely will be grouped into the same cluster. This is intuitively nontrival as the users locating together are likely to have similar channel conditions and thus can be efficiently served by a hovering UAV.
As a first attempt, treating the users' position information as feature data, \cite{cui-clustering} proposed a k-means based unsupervised clustering algorithm to solve the user clustering problem. However,
the k-means based approach is sensitive to the initial cluster centroids selection and improper initial cluster centroids may result in undesirable user clustering outcomes.
Therefore, it is of great interests to design a more robust machine learning based approach to
efficiently solve the user clustering problem. Under imperfect CSI assumption, the worse-case
robust beamforming design problem, with signal-to-interference-noise-ratio (SINR)
and SIC constraints, is also investigated. Different to \cite{pan-uav,new-uav}, which considered either  the single-cell or single-antenna scenario, this work focus on a multi-antenna scenario in the multi-cell interference channel. The coupling effects of the beamformers, both from inter- and intra-cluster, make the formulated problem nonconvex and challenging to solve. Moreover, the imperfect CSI assumption make the considered problem much more complicated, due to the fact that each of the worst-case SINR or SIC constraints corresponds to an infinite number of nonconvex constraints. For more practical applications, we further investigate the algorithm to solve the robust beamforming design problem in a decentralized fashion. 
The contributions of this work are summarized as follows: 
\begin{itemize}
	\item We formulate the joint user clustering and robust beamforming problem in a multi-antenna
	UAV-NOMA system under imperfect CSI assumptions. For computation-efficiency consideration, the original problem is decoupled into two subproblems, i.e., the user clustering problem and the robust beamforming problem. To solve the user clustering problem
	efficiently, we propose to use a k-means++ based unsupervised learning approach, which
	consists of a careful initial cluster centroids selection process and a standard k-
	means based user clustering process.
	\item To attain some useful insights on solving the robust beamforming design problem, we first consider a special case with perfect CSI.  By equivalently transforming the associated problem into a second-order cone programming, the relevant problem is solved optimally.
	Then, we focus on the problem in the general case with imperfect CSI. To simplify the corresponding problem, we first use the semidefinite relaxation (SDR) based method to transform the quadratic terms respect to beamformers into linear ones, and then, the S-lemma is invoked to dealt with the infinitely many constraints caused by the imperfect CSI. 
	By omitting the rank-one constraints, the reformulated problem refers to the semidefinite programming (SDP) problem, which is convex and can be efficiently solved by the existing optimization tools. 
	Further, to gain more insights on the proposed SDR based algorithm, a sufficient condition, under which the rank-one optimality of the obtained solution can be guaranteed, is provided and the rank-one optimality of the obtained solution is theoretically proved.
	\item For practical applications, we investigate the decentralized approach to solve the robust beamforming design problem. By equivalent reformulations, the constraints set is decoupled
	into several independent subsets, each of which is solely related to a single UAV. Then, the alternating direction method of multipliers (ADMM) is applied to solve the reformulated problem efficiently.
\end{itemize}
Simulation results are presented to show the efficiency of the proposed transmission scheme and
algorithms. The rest of the paper are organized as follow. In Section \ref{sec:system-model}, the considered system model and the transmission power minimization problem is introduced.
Section \ref{sec:clustering} presents the k-means++ based algorithm to solve the user clustering problem. In Section \ref{sec:robust-beamforming}, the centralized robust beamforming design is investigated. While the decentralized robust beamforming design is studied in Section \ref{sec:admm}. Simulation results and conclusions are given in Sections \ref{sec:simulation} and \ref{sec:conclusion} respectively.

\textit{Notations}: Column vectors and matrices are denoted by boldfaced lowercase and uppercase letters, e.g., ${\bf x}$ and $\bf X$; $\mathbb{R}^n$, $\mathbb{C}^n$ and $\mathbb{H}^n$ stand for the sets of $n$-dimensional real and complex vectors and complex Hermitian matrices, respectively.
$\bf{I}^n$ denotes the $n \times n$  identity matrix, and $\bf 0 (1)$ denotes an all-zero (one) vector (matrix) with appropriate dimension. ${\bm{\mathcal{W}}}^+$ indicates that the element of ${\bf W}$ are non-negative. The superscripts $(\cdot)^{T}$,  $(\cdot)^{H}$ and  $(\cdot)^{\dagger}$  describe the transpose, (Hermitian) conjugate transpose and pseudo-inverse operations, respectively. and $\rank(\bf X)$ and $\tr(\bf X)$ represent the rank and trace of matrix, respectively. ${\bf X} \succeq (\succ) ~{\bf 0}$ means that matrix $\bf X$ is positive semidefinite (positive definite). $||\bf x||_2$ denotes the Euclidean norm of vector ${\bf x}$. $\mathbb{E}\{\cdot\}$ denotes the statistical expectation. 

\section{System Model and Problem Formulation} \label{sec:system-model}
Consider a downlink communication network which consists of $M$ multi-antenna UAVs and $K$ single-antenna users. Let $\mathcal{M} \triangleq \{1,2,...,M\}$ and $\mathcal{K} \triangleq \{1,2,...,K\}$ denote the index sets of UAVs and users, respectively.
According to their locations, the users are grouped into $M$ non-overlapping user clusters, i.e., $\mathcal{C}_n \cap \mathcal{C}_{\hat{n}} = \emptyset, \forall n \neq \hat{n}$ where $\mathcal{C}_n \triangleq \{{\rm U}_{n1},{\rm U}_{n2},...,{\rm U}_{nN_n}\}$ denotes the user set of the $n$-th cluster and  $N_n$ is the number of users in the $n$-th cluster, satisfying $\sum_{n=1}^{M} N_n = K$. 
Each cluster is served by a hovering UAV. Without loss of generality, the users are assumed to 
locate in the same plane. Let ${\bf p}_{nk}^{\rm user} = [\bar{x}_{nk}, \bar{y}_{nk}, 0]^T$ and ${\bf p}_{n}^{\rm uav} =[{x_n^c, y_n^c, h_n}]^T$ denote the coordinate vectors of user $k$ in the $n$-th cluster and UAV $n$, respectively. In particular, $h_n$ is the flying height of UAV $n$ and $x_n^c$, $y_n^c$ denote the coordinate of the centroid of the $n$-th cluster, which can be calculated by
\begin{align}
x_n^c = \frac{1}{N_n}\sum_{k=1}^{N_n} \bar{x}_{nk}, ~y_n^c =  \frac{1}{N_n}\sum_{k=1}^{N_n} \bar{y}_{nk}, \forall n\in\mathcal{M}.
\end{align}

\subsection{User Location Model}
 Unlike the conventional assumption in the previous work that the users are uniformly distributed in the network \cite{Ding-2014-SPL,fang-jsac-2017}, we focus on the scenario where the users' locations are physically correlated in this work. This scenario characterizes the case where the system contains some hotspots, such as sports bar or lecture hall, etc \cite{hetnet}. In this case, the high maneuverability of UAV can be exploited to provide specific services for each hotspot. The Possion Cluster Process (PCP) is used to model the users' distribution. Specifically, the PCP can be model by \cite{pcp}
 \begin{align}
 \phi = \cup_{{\bf z} \in \phi_p} {\mathcal{\bf z}} + \mathcal{B}^{\mathcal{\bf z}} 
 \end{align}
 where $\phi_p$ denotes the process of the parent points  and $\mathcal{B}^{\mathcal{\bf z}}$ denotes the off-spring points process associated with the cluster center ${\bf z}$. We assume that the parent points are uniformly distributed in the network and the off-spring points in the $n$-th cluster also follows the uniform distribution in a circular range, with radius of $r_{n}$ , around the cluster center. 

\subsection{Channel Model}
Note that the users in a cluster are severed by a UAV hovering above them, thus there is a high possibility that there exits a line-of-sight (LoS) communication link between the UAV  and its home user. Therefore, the channel between the UAV and its home user is modelled by the Rician fading channels. While, due to the blockages of the tall-buildings or trees, it is possible that no direct communication link exists between the UAV  and its neighbouring users. Hence, the Rayleigh fading channel is more suitable for modelling the channels between UAV and its neighbouring users.

Let ${\bf h}^H_{mnk}\in \Cset^{N_t}$ denote the channel from UAV $m$ to U$_{nk}$. As discussed previously that the UAVs inevitably suffer from CSI errors in practice. Thus, imperfect CSI model is considered in this work. Let $\tilde{{\bf h}}_{mnk}\in \Cset^{N_t}$ denote the pre-assumed CSI at the UAV $m$ for U$_{nk}$. Then, the real CSI between UAV $m$ and U$_{nk}$ is given by
\begin{align}
{\bf h}_{mnk} = \tilde{{\bf h}}_{mnk} + {\bf e}_{mnk}, \forall m,n \in \mathcal{M},k \in \mathcal{N}_n, 
\end{align}
where ${\bf e}_{mnk}$ is the bounded CSI error associated with ${\bf h}_{mnk}$. In particular, the bounded CSI error can be modelled by
\begin{align}
{\bf e}^H_{mnk} {\bf Q}_{mnk} {\bf e}_{mnk} \leq 1, \forall m,n \in \mathcal{M},k\in \mathcal{N}_n,
\end{align}
where ${\bf Q} \in \mathbb{H}^{N_t}$ determines the range and shape of the CSI error. For instance, ${\bf Q} = \frac{1}{\epsilon^2}{\bf I}_{N_t} $ characterizes the popular spherical error model $||{\bf e}_{mnk}||^2 \leq \epsilon^2$. 

\subsection{Transmission Model and Problem Formulation}
For spectral efficiency consideration, the NOMA protocol is applied to each user cluster. Follow the rational of NOMA, the signals of users in cluster $n$ are combined by using the superposition coding technique at UAV $n$, then the users with stronger channel conditions will first remove the signals for the users with weaker channel conditions by invoking the SIC technique.
Let  $s_{nk} \in \Cset$ denote the signal for user $k$ in cluster $n$ with $\mathbb{E}\{|s_{nk}|^2\} = 1$. So, after superposition coding, the transmit signal of UAV $n$ is given by
\begin{align}
{\bf s}_{n} = \sum_{k=1}^{N_n} {\bf w}_{nk} s_{nk}, \forall n \in \mathcal{M},
\end{align}
where  ${\bf w}_{nk}$ is the beamformer for U$_{nk}$.
The received signal at U$_{nk}$ is give by
\begin{align} \label{eq:received_signal}
y_{nk} = &{\bf h}^H_{nnk}{\bf w}_{nk} s_{nk} + {\bf h}^H_{nnk} \sum_{i = 1, i \neq k}^{N_n}{\bf w}_{ni} s_{ni} + \sum_{m=1,m\neq n}^M {\bf h}^H_{mnk} \sum_{i=1}^{N_m}{\bf w}_{mi} s_{mi} + z_{nk},
\end{align}
where $z_{nk} \in \Cset$ is the received additive white Gaussian noise at U$_{nk}$ with zero mean and variance $\sigma_{nk}^2$. The first item in \eqref{eq:received_signal} denotes the desired signal of U$_{nk}$, the second and third items in \eqref{eq:received_signal} denote the intra-cell and inter-cell interference, respectively.


Without loss of generality,  we assume that the users, in each cluster, are ordered by their channel gains in a descending manner, i.e., $|{\bf h}_{mn1}|^2\geq|{\bf h}_{mn2}|^2\geq\cdots|{\bf h}_{mnN_n}|^2$. Thus, according to the principle of NOMA, U$_{nk}$ would first remove the information, $s_{nj}$, for U$_{nj}$ for $j>k$ by using SIC, and then decoding its own information. Based on the above model, the SINR for decoding $s_{nj}$, $\forall j>k$, at U$_{nk}$ in the SIC process is given by
\begin{align}
&\SINR_{nk}^{s_{nj}}\left(\{{\bf w}_{nk}\}_{\forall n,k},\{{\bf h}_{mnk}\}_{\forall m,n,k}\right) = \notag\\
&~~~~~~ \frac{|{\bf h}^H_{nnk}{\bf w}_{nj}|^2}{\sum_{i=1}^{j-1}|{\bf h}^H_{nnk} {\bf w}_{ni}|^2 + \sum_{m=1,m\neq n}^M \sum_{i=1}^{N_m}|{\bf h}^H_{mnk}{\bf w}_{mi}|^2 + \sigma_{nk}^2}, \forall {\bf e}^H_{mnk} {\bf Q}_{mnk} {\bf e}_{mnk} \leq 1,
\end{align}

After $s_{nj}$ is removed from the received signal, the SINR at U$_{nk}$ for decoding $s_{nk}$ is given by
\begin{align}
&\SINR_{nk}^{s_{nk}}\left(\{{\bf w}_{nk}\}_{\forall n,k},\{{\bf h}_{mnk}\}_{\forall m,n,k}\right) = \notag\\
& ~~~~~~\frac{|{\bf h}^H_{nnk}{\bf w}_{nk}|^2}{\sum_{i = 1}^{k-1}|{\bf h}^H_{nnk} {\bf w}_{ni}|^2 + \sum_{m=1,m\neq n}^M \sum_{i=1}^{N_m}|{\bf h}^H_{mnk}{\bf w}_{mi}|^2 + \sigma_{nk}^2}, \forall {\bf e}^H_{mnk} {\bf Q}_{mnk} {\bf e}_{mnk} \leq 1,
\end{align}

Based on the above model, the energy-efficient joint user-clustering and robust beamforming design problem can be formulated as 
\begin{subequations}\label{p:p1}
	\begin{align}
&\!\!\!\!\!\!\!\!\!\!\!\!\!\!\!\min_{\{{\mathcal{C}_n}\}_{n=1}^{M}, \{{\bf w}_{nk}\}_{n,k}}~ ~~\sum_{n=1}^{M} \sum_{k=1}^{N_n} ||{\bf w}_{nk}||^2\\
    \st~& \mathcal{C}_n \cap \mathcal{C}_{n'} = \emptyset, \forall n \neq n' \label{eq:p1_clutering}\\
	&\SINR_{nk}^{s_{nj}}\!\left(\{{\bf w}_{nk}\}_{n,k},\{{\bf h}_{mnk}\}_{m,n,k}\right)\! \!\geq\! \gamma_{nj}, \forall {\bf e}^H_{mnk} {\bf Q}_{mnk} {\bf e}_{mnk} \!\leq\! 1,m,n\!\in\! \mathcal{M},k\!\in\! \mathcal{C}_n, j\!>\!k, \label{eq:p1_sic}\\
	& \SINR_{nk}^{s_{nk}}\left(\{{\bf w}_{nk}\}_{n,k},\{{\bf h}_{mnk}\}_{m,n,k}\right) \geq \gamma_{nk}, \forall {\bf e}^H_{mnk} {\bf Q}_{mnk} {\bf e}_{mnk} \leq 1, m,n \in \mathcal{M}, k\in \mathcal{C}_n, \label{eq:p1_qos}\\
	& \sum_{k=1}^{N_n} ||{\bf w}_{nk}||^2 \leq P_{\max},  \forall n \in \mathcal{M},
	\end{align}
\end{subequations}
where $P_{\max}$ is the transmission power budget of each UAV. \eqref{eq:p1_clutering} represents that each user will be uniquely assigned into one cluster; \eqref{eq:p1_sic} guarantees the success of SIC procedure at each user and the Quality-of-Service requirement for each user is given in \eqref{eq:p1_qos}. 

It is not difficult to verify that problem \eqref{p:p1} is a nonconvex optimization problem due to the coupling of the quadratic beamforming vectors and also the channel uncertainty. More precisely, problem \eqref{p:p1} is in fact an NP-hard mixed integer programming problem and, thus, is unsolvable within polynomial time. To efficiently produce a high quality solution of problem \eqref{p:p1}, similar to \cite{di-user-pairing,cui-user-pairing,liang2017,xu-tcom-harq}, we will decouple it into two subproblems, i.e., the user clustering problem and the robust beamforming design problem. However, unlike \cite{di-user-pairing,cui-user-pairing,liang2017,xu-tcom-harq}, where the user paring problem is solved by the traditional optimization methods, in this work, we will use an unsupervised clustering based approach to efficiently produce a high-quality clustering outcome with much lower computational complexity. In the next section, the details of the k-means++ based unsupervised user clustering algorithm will be discussed.  

\section{A K-Means++ based Approach for User Clustering}\label{sec:clustering}
Notice that the users are spatially located in several clusters by following the PCP. It motivates us to design an efficient user clustering algorithm by exploring the distribution information of the users' positions.  
As mentioned previously, to benefit the advantages of unsupervised learning and the users' position information, \cite{cui-clustering} proposed a k-means based algorithm to perform fast user clustering, but with the curse of the sensitivity to the initial centroids selection.
In view of this, a new unsupervised learning based algorithm will be proposed in this work. To this end, by utilizing the users' position information, we first rewrite the user clustering problem as the following Euclidean distance-oriented optimization problem
\begin{subequations}\label{p:clustering}
	\begin{align}
	\min_{{\bf C},{\bf X}} ~&||{\bf P} - {\bf C} {\bf X}||_2^2\\
	\st ~&{\bf 1}^T{\bf X}_k = 1, \forall k\in\mathcal{C}_n,\label{eq:assignment}\\
	&[{\bf X}]_{nk} \in \{0,1\}, \forall n \in \mathcal{M},k\in\mathcal{C}_n, \\
	& {\bf C} \in {\bm{\mathcal{C}}}^+,
	\end{align}
\end{subequations}
where ${\bf P} \in \Rset^{3\times K}$ collects the positions of all users, ${\bf C} \in \Rset^{3\times M}$ represents the centroids of the $M$ user clusters and ${\bf X} \in \Rset^{M\times K}$ indicates the cluster assignment of users. Specifically, $[{\bf X}]_{nk} = 1$ implies that user $k$ is assigned to the $n$-th cluster, $[{\bf X}]_{nk} = 0$ otherwise. Constraint \eqref{eq:assignment} indicates that each user can only be grouped into one cluster.

Notice that the constraint sets on ${\bf C}$ and ${\bf X}$ are decomposable. Thus, problem \eqref{p:clustering} can be efficiently solved by the alternating optimization (AO) based algorithm. In particular, by using the AO based algorithm, the user clustering outcome and the centroids of the clusters would be alternatively updated until the clustering converges. This is exactly the well-known k-means algorithm \cite{k-means}, which has been widely used to solve the data clustering problem, e.g., \cite{cui-clustering}. However, due to the non-convex integer constraint in \eqref{eq:assignment}, the k-means algorithm is sensitive to the initial selection of the cluster centroids and may not always yield satisfactory clustering performance \cite{k-means++}. 
 In view of this, the k-means++ algorithm, which is an improvement of k-means algorithm, will be used to solve the clustering problem in this work. 
 
The ingredients of the k-means++ method are two-folds: one is the initial cluster centroids selection process; the other is the standard k-means method to find the final clustering outcome.
The basic idea of the initial cluster centroids selection is as follow. The system first randomly selects a user as the first centroid. Secondly, the system needs to compute the distances from all the other users to this centroid, denoted by $d_{k}, \forall k\in\mathcal{C}_n$. Then, user $k$ will be selected as the second cluster centroid with probability $\mathcal{P}_k = \frac{d_k^2}{\sum_{k=1}^Kd_k^2}$. Thirdly, recompute the distances from all users to these two selected centroids. Then, let $d_k = \min \{d_{1,k}, d_{2,k}\}$ where $d_{1,k}$ and $d_{2,k}$ denote the distances from user $k$ to the first and second cluster centroid. Again, choose user $k$ as the third cluster centroid with probability $\mathcal{P}_k$. Repeat the above steps until all the $K$ centroids are selected. Finally, standard k-means algorithm will be applied to solve the user clustering problem based on the chosen cluster centroids. The detailed steps of the k-means++ based user clustering algorithm is summarized in Algorithm \ref{alg:k-means++}.

\begin{algorithm}[!tb] \smaller[1]
	\caption{K-means++ based algorithm for user clustering}\label{alg:k-means++}
	\begin{algorithmic}[1]
		\STATE {{\bf Given} the users' positions ${\bf P}$ and the number of user clusters $M$.} \\
		\vspace{1mm}
		{\setlength\parindent{-1.1em} \bf Phase I : Initial cluster centroids selection} 	\vspace{1mm}
		\STATE {Choose a user from $\mathcal{K}$ as the first centroid ${\bf c}_1$ randomly.}
		\FOR {$n = 2:M$}
		\STATE {Compute the Euclidean distances, ${\bf dist}({\bf c}_{\ell},{\bf P}_k), \forall  1\leq \ell \leq n-1, k\in\mathcal{C}_n$, between all users and the chosen centroids $\{{\bf c}_{\ell}\}_{\ell=1}^{n-1}$, and denote the distances by $\{{\bf d}_{1k},...,{\bf d}_{\ell k}\}, \forall k\in\mathcal{C}_n$.}
		\STATE {Find the smallest distance of users to the chosen centroids $d_k = \min \{{\bf d}_{1k},...,{\bf d}_{\ell k}\}, \forall k\in\mathcal{C}_n$, and store in ${\bf d} = [d_1,..., d_K]$.}
		\STATE {Choose use $k$ as centroid ${\bf c}_n$ with probability $\frac{d_k^2}{\sum_{k=1}^Kd_k^2}$}
		\ENDFOR\\
			\vspace{1mm}
		{\setlength\parindent{-1.1em} {\bf Phase II : K-means based user clustering} }	\vspace{1mm}
		\WHILE {the clustering does not converge}
		\STATE {Compute the distances between all users and centroids, ${\bf dist}({\bf c}_n,{\bf P}_k), \forall n\in\mathcal{M}, k\in\mathcal{C}_n$.}
		\STATE {Include user $k, \forall k\in\mathcal{C}_n$, into cluster $n$, $\forall n \in \mathcal{M}$ with the smallest distance.}
		\STATE {Update the new centroids with:
		\begin{align}
		{\bf c}_n = \frac{1}{N_n} \sum_{k=1}^{N_n} {\bf P}_k, \forall n \in \mathcal{M}.
		\end{align}}
		\ENDWHILE
		\STATE {{\bf{Output :}} The clustering outcome $\mathcal{C}_n, \forall n \in \mathcal{M}$}
	\end{algorithmic}
\end{algorithm}

From Algorithm \ref{alg:k-means++}, we can see that, one important trick of the initial cluster centroids selection is to choose the user that has a lager distance to the chosen centroids as the next centroid with a higher probability. This is intuitively reasonable, as larger distance between centroids results in a more robust user clustering outcome. Also, note that the k-means++ algorithm chooses the next centroid with a probability, instead of choosing the user with the largest distance to the chosen centroids as the next centroid directly.
This is to combat the effects from the {noise point}\footnote{The noise point in data clustering means that this point is far from all the other data points in the data set.}.

It is important to point out that, at the first glance, the computational complexity of the k-means++ method would be higher than that of the standard k-means method, as an additional initial cluster centroids selection process has been added. However, generally, the convergence behaviour of the k-means++ can perform better than the k-means method thanks to the careful selection of the initial cluster centroid selection. Although it is difficult to quantify theoretically, the performance efficacy, both in accuracy and speed, of the  k-means++ algorithm has been verified on various data sets, compared to the standard k-means algorithm, see \cite[section 6]{k-means++} for more details.

\section{Solve the Robust Beamforming Design Problem} \label{sec:robust-beamforming}
Once the user clustering is determined, problem \eqref{p:p1} boils down to a pure robust beamforming design problem. In this section, we will first optimally solve a special case of problem \eqref{p:p1} under the perfect CSI assumption to gain some insights on the potential difficulty in solving problem \eqref{p:p1}. Then, we investigate the robust beamforming design present in problem \eqref{p:p1} and an SDR-based algorithm is proposed to produce a suboptimal solution in an efficient way. Finally, the condition, under which the proposed SDR-based algorithm can generate an optimal solution, is studied.

\subsection{Optimal Design for the Case under Perfect CSI Assumption} \label{subsec:non-robust}
Recall problem \eqref{p:p1} under the assumption of perfect CSI as 
\begin{subequations}\label{p:p2}
	\begin{align}
	\min_{\{{\bf w}_{nk}\}_{\forall n,k}}&~ \sum_{n=1}^{M} \sum_{k=1}^{N_n} ||{\bf w}_{nk}||^2\\
	\st~~&\frac{1}{\gamma_{nj}}|{\bf \hat{h}}^H_{nnk}{\bf w}_{nj}|^2 \geq \sum_{i=1}^{j-1}|{\bf \hat{h}}^H_{nnk} {\bf w}_{ni}|^2 + \sum_{m=1,m\neq n}^M \sum_{i=1}^{N_m}|{\bf \hat{h}}^H_{mnk}{\bf w}_{mi}|^2 + \sigma_{nk}^2, \forall m,n\in \mathcal{M}, k \in \mathcal{C}_n,\\
	& \frac{1}{\gamma_{nk}}|{\bf \hat{h}}^H_{nnk}{\bf w}_{nk}|^2 \geq \sum_{i = 1}^{ k-1}|{\bf \hat{h}}^H_{nnk} {\bf w}_{ni}|^2 + \sum_{m=1,m\neq n}^M \sum_{i=1}^{N_m}|{\bf \hat{h}}^H_{mnk}{\bf w}_{mi}|^2 + \sigma_{nk}^2 , \forall m,n \in \mathcal{M}, k\in \mathcal{C}_n,\\
	& \sum_{k=1}^{N_n} ||{\bf w}_{nk}||^2 \leq P_{\max},  \forall n \in \mathcal{M}.
	\end{align}
\end{subequations}

By epigraph reformulation, problem \eqref{p:p2} can be equivalently reformulated as
\begin{subequations}\label{p:socp}
	\begin{align}
	&\min_{t, \{{\bf w}_{nk}\}_{\forall n,k}}  ~~t \\
	\st &\left(\sum_{n=1}^{M} \sum_{k=1}^{N_n} ||{\bf w}_{nk}||^2\right)^{\frac{1}{2}} \leq t, \\
	&\left(\sum_{i=1}^{j-1}|{\bf \hat{h}}^H_{nnk} {\bf w}_{ni}|^2 \!+\!\!\!\! \sum_{m=1,m\neq n}^M \sum_{i=1}^{N_m}|{\bf \hat{h}}^H_{mnk}{\bf w}_{mi}|^2 \!+\! \sigma_{nk}^2\right)^{\frac{1}{2}}\!\! \!\leq\!\! \frac{1}{\sqrt{\gamma_{nj}}}{\bf \hat{h}}^H_{nnk}{\bf w}_{nj}, \forall m,n \!\in\!\mathcal{M}, k \!\in\! \mathcal{C}_n,j\!>\!k,\\
	& \left(\sum_{i = 1}^{k-1}|{\bf \hat{h}}^H_{nnk} {\bf w}_{ni}|^2 \!\!+\!\!\!\! \sum_{m=1,m\neq n}^M \sum_{i=1}^{N_m}|{\bf \hat{h}}^H_{mnk}{\bf w}_{mi}|^2 \!+\! \sigma_{nk}^2\right)^{\frac{1}{2}} \!\!\!\leq\!\! \frac{1}{\sqrt{\gamma_{nk}}}{\bf \hat{h}}^H_{nnk}{\bf w}_{nk} , \forall m,n\!\in\!\mathcal{M}, k \!\in\! \mathcal{C}_n,\\
	& \left( \sum_{k=1}^{N_n} ||{\bf w}_{nk}||^2 \right)^{\frac{1}{2}}\leq \sqrt{P_{\max}},  \forall n \in \mathcal{M},
	\end{align}
\end{subequations}
which is a second-order cone programming and thus can be efficiently solved by the interior point based solver, e.g., {\tt CVX} \cite{cvx}. Now, one can realize that the challenge in solving the robust beamforming design problem is rather than the coupling effect of the quadratic beamformers, the channel uncertainty instead. In the next subsection, we will focus on the robust beamforming design problem.

\subsection{Suboptimal Design for the General Cases}
In this section, we first propose to use the SDR method to relax the quadratic terms related to the beamforming vectors to linear ones. Then, we handle the obstacle bringing by the infinitely many SINR constraints. Finally, we also provide a sufficient condition under which the SDR will be tight.
\vspace{2mm}

\noindent {\bf B.1. SDR-Based Suboptimal Design : } 
To apply the SDR method, we first introduce a set of rank-one matrix ${\bf W}_{n,k} = {\bf w}_{nk}{\bf w}_{nk}^H, \forall n,k$. Then, by ignoring the rank-one constraint on the matrix, the robust beamforming problem can be relaxed as
\begin{subequations}\label{p:sdr_robust}
	\begin{align}
	\min_{\{{\bf W}_{nk}\}_{\forall n,k}}~ &\sum_{n=1}^{M} \sum_{k=1}^{N_n} \tr({\bf W}_{nk})\\
	\st~~~&{\bf h}^H_{nnk} \left(\frac{1}{\gamma_{nj}}{\bf W}_{nj} - \sum_{i=1}^{j-1} {\bf W}_{nk}\right){\bf h}_{nnk} \geq \sum_{m=1,m\neq n}^M {\bf h}^H_{mnk} \left(\sum_{i=1}^{N_m} {\bf W}_{mi}\right) {\bf h}_{mnk}  + \sigma_{nk}^2, \notag\\
	&\quad\quad\quad\quad\quad\quad\quad\quad\quad\quad\quad\quad\forall {\bf e}^H_{mnk} {\bf Q}_{mnk} {\bf e}_{mnk} \leq 1, m,n \in\mathcal{M}, k \in \mathcal{C}_n,j>k,\label{eq:sdr_sic}\\
	& {\bf h}^H_{nnk} \left(\frac{1}{\gamma_{nk}}{\bf W}_{nk} - \sum_{i=1}^{k-1} {\bf W}_{ni}\right){\bf h}_{nnk} \geq \sum_{m=1,m\neq n}^M {\bf h}^H_{mnk} \left(\sum_{i=1}^{N_m} {\bf W}_{mi}\right) {\bf h}_{mnk}  + \sigma_{nk}^2, \notag\\
	&~~\quad\quad\quad\quad\quad\quad\quad\quad\quad\quad\quad\quad\quad\quad\forall {\bf e}^H_{mnk} {\bf Q}_{mnk} {\bf e}_{mnk} \leq 1, m,n\in\mathcal{M}, k\in \mathcal{C}_n, \label{eq:sdr_qos}\\
	& \sum_{k=1}^{N_n} \tr({\bf W}_{nk}) \leq P_{\max}, \forall n\in\mathcal{M}, \\
	& {\bf W}_{nk} \succeq {\bf 0}, \forall n \in \mathcal{M}, k \in \mathcal{C}_n,
	\end{align}
\end{subequations}
which is a convex problem as the objective function and constraints are linear. However, it is still computationally intractable due to the infinite number of constraints. Next, to make problem \eqref{p:sdr_robust} tractable, we propose the following lemma. 

\begin{lem}\label{lem:s-lemma}
	The infinitely many constraints in \eqref{eq:sdr_sic} and \eqref{eq:sdr_qos} can be equivalently recast into the following finite number of linear matrix inequalities:
	\begin{subequations}\label{eq:lemma1}
		\begin{align}
		& {\bf \Phi}_{nj}\left(\{{\bf W}_{ni}\}_{i=1}^j,\{\theta_{mnk}\}_m,\lambda_{mnk}\right) \triangleq \notag\\
		&\left[
		\begin{matrix}
		{\bf A}_{nj} \!+\!\lambda_{nnk}{\bf Q}_{nnk} & {\bf A}_{nj}\hat{\bf h}_{nnk}\\
		\hat{\bf h}_{nnk}^H {\bf A}_{nj} & \hat{\bf h}_{nnk}^H {\bf A}_{nj}\hat{\bf h}_{nnk} -\!\! \sum\limits_{m=1,m\neq n}^M \!\!\theta_{mnk} \!-\!\sigma_{nk}^2 \!-\!\lambda_{nnk}
		\end{matrix}
		\right] \!\succeq\! {\bf 0}, \forall m,n \in\mathcal{M}, j,k \!\in\! \mathcal{C}_n, j\!>\!k,\\
		&{\bf \Psi}_{nk}\left(\{{\bf W}_{ni}\}_{i=1}^k,\{\theta_{mnk}\}_m,\lambda_{mnk}\right) \triangleq \notag\\
		& \left[
		\begin{matrix}
		{\bf B}_{nk} \!+\!\lambda_{nnk}{\bf Q}_{nnk} & {\bf B}_{nk}\hat{\bf h}_{nnk}\\
		\hat{\bf h}_{nnk}^H {\bf B}_{nk} & \hat{\bf h}_{nnk}^H {\bf B}_{nk}\hat{\bf h}_{nnk} \!-\!\! \sum\limits_{m=1,m\neq n}^M \!\!\theta_{mnk} \!-\!\sigma_{nk}^2 \!-\!\lambda_{nnk}
		\end{matrix}
		\right] \!\succeq\! {\bf 0},\!\forall m,n \!\in\! \mathcal{M}, k \!\in\! \mathcal{C}_n, \\
		&{\bf \Omega}_{mnk}\left(\{{\bf W}_{ni}\}_{i=1}^{N_n},\theta_{mnk},\lambda_{mnk}\right) \triangleq \notag\\
		& ~~~~\left[
		\begin{matrix}
		{\bf -C}_{m} +\lambda_{mnk}{\bf Q}_{mnk} & -{\bf C}_{m}\hat{\bf h}_{mnk}\\
		-\hat{\bf h}_{mnk}^H {\bf C}_{m} & -\hat{\bf h}_{mnk}^H {\bf C}_{m}\hat{\bf h}_{mnk} + \theta_{mnk} + \lambda_{mnk}
		\end{matrix}
		\right] \succeq {\bf 0},\forall m,n \in \mathcal{M}, k\in\mathcal{C}_n,
		\end{align}
	\end{subequations}
where $\lambda_{mnk} > 0, \forall m,n \in \mathcal{M}, k\in\mathcal{C}_n$ are auxiliary variables and 
\begin{align*}
	{\bf A}_{nj} &= \frac{1}{\gamma_{nj}}{\bf W}_{nj} - \sum_{i=1}^{j-1} {\bf W}_{nk}, \\
	{\bf B}_{nk} &= \frac{1}{\gamma_{nk}}{\bf W}_{nk} - \sum_{i=1}^{k-1} {\bf W}_{ni}, \\
	{\bf C}_m &= \sum_{i=1}^{N_m} {\bf W}_{mi},\\
	 \theta_{mnk} &= \max_{\forall {\bf e}^H_{mnk} {\bf Q}_{mnk} {\bf e}_{mnk} \leq 1,} {\bf h}^H_{mnk} \left(\sum_{i=1}^{N_m} {\bf W}_{mi}\right) {\bf h}_{mnk}.
\end{align*}
\end{lem}
\begin{IEEEproof}
	The key idea to prove Lemma \ref{lem:s-lemma} is using S-lemma to handle the infinitely many constraints. The detailed proof is relegated to Appendix \ref{apdx:s-lemma}.
	\end{IEEEproof}

Based on Lemma \ref{lem:s-lemma}, the SDP problem \eqref{p:sdr_robust} can be equivalently reformulated as 
	\begin{subequations}\label{p:sdp}
	\begin{align}
	\min_{\{{\bf W}_{nk}\},\{\theta_{mnk}\},\{\lambda_{mnk}\}} &\sum_{n=1}^{M} \sum_{k=1}^{N_n} \tr({\bf W}_{nk})\\
	\st ~~~& {\bf \Phi}_{nj}\left(\{{\bf W}_{ni}\}_{i=1}^j,\{\theta_{mnk}\}_m,\lambda_{nnk}\right) \succeq {\bf 0},\forall m,n \in \mathcal{M}, j,k\in\mathcal{C}_n, j>k,\\
	&{\bf \Psi}_{nk}\left(\{{\bf W}_{ni}\}_{i=1}^k,\{\theta_{mnk}\}_m,\lambda_{nnk}\right) \succeq {\bf 0},\forall m,n \in \mathcal{M}, k\in\mathcal{C}_n, \\
	&{\bf \Omega}_{mnk}\left(\{{\bf W}_{ni}\}_{i=1}^{N_n},\theta_{mnk},\lambda_{mnk}\right) \succeq {\bf 0},\forall m,n \in \mathcal{M}, k\in\mathcal{C}_n, \\
	& \sum_{k=1}^{N_n} \tr({\bf W}_{nk}) \leq P_{\max}, \forall n\in\mathcal{M}, \\
	& {\bf W}_{nk} \succeq {\bf 0}, \lambda_{mnk} \geq 0, \forall n \in \mathcal{M}, k\in\mathcal{C}_n,
	\end{align}
\end{subequations}
which is an SDP and thus can be efficiently solved by {\tt{CVX}}. 

Remind that problem \eqref{p:sdr_robust} is a relaxed version of the original robust beamforming design problem by ignoring the rank-one constraints. Thus, one important issue in solving problem \eqref{p:sdr_robust} is to verify whether the obtained matrices from solving problem \eqref{p:sdp} are rank-one. If it is true, then the optimal beamforming vectors can be obtained by simply applying singular value decomposition to the obtained matrices. Hence, it is interesting to explore the conditions under which solving problem \eqref{p:sdp} can produce rank-one solutions.\vspace{2mm}

\noindent {\bf B.2. Rank-One Optimality Analysis: } The following Lemma provides a condition that can guarantee the rank optimality of problem \eqref{p:sdp}.

\begin{lem}\label{lem:rank-one} Suppose that problem \eqref{p:sdp} is feasible,
	the rank-one optimality can be guaranteed if ${\bf Q}_{nnk}= \infty {\bf I}_{N_t}$ for all $n, k$, i.e., no intra-cell CSI error.
\end{lem}
\begin{IEEEproof}
	The proof is relegated to Appendix \ref{app:rank}.
\end{IEEEproof}

We note that, for a general setup, the solutions of problem \eqref{p:sdp} {\textit{may not}} be rank-one. In these cases, one can resort to Gaussian randomization method to produce approximated beamforming vectors based on the obtained non-rank-one solutions \cite{Luo-2010}.

\section{Decentralized Beamforming Design via ADMM} \label{sec:admm}
In the previous section, we propose to solve the robust beamforming design problem \eqref{p:sdp} by using SDR in a centralized fashion. However, solving problem \eqref{p:sdp}, in this way, is based on the assumption that there is central controller to collect the entire CSI of the UAVs. In this case, the signalling overhead would become heavier as the increase of the number of users. Moreover, such a central controller may not be always available in practice. Therefore, it is of great importances to investigate the approaches that solve problem \eqref{p:sdp} in a decentralized way. In view of this, we propose to use the well-known ADMM method to solve problem \eqref{p:sdp} in a decentralized fashion. However, applying ADMM method to problem \eqref{p:sdp} is not straightforward. Fortunately, by equivalent reformulations, problem \eqref{p:sdp} can be transformed to an appropriate structure under which the ADMM method can be applied. To this end, we first define the following auxiliary variables:
\begin{align}
p_n = \sum_{k=1}^{N_n} {\bf W}_{nk},~\Theta_{nk} = \sum_{m=1,m\neq n}^{N_m} \theta_{mnk}, \forall n \in \mathcal{M}, k\in\mathcal{C}_n,
\end{align}
where $p_n$ is the transmission power of UAV $n$, and $\Theta_{mnk}$ is the received inter-cell interference power from the other UAVs to U$_{nk}$ in the system. Then, problem \eqref{p:sdp} can be reformulated as
\begin{subequations} \label{eq:admm_p1}
	\begin{align}
	\min_{\substack{\{{\bf W}_{nk}\succeq{\bf 0}\},\{p_n\},\\\{\theta_{mnk}\},\{\Theta_{mnk}\},\{\lambda_{mnk} \geq 0\}}}~&\sum_{n=1}^{M} p_n \\
	\st ~~~& {\bf \Phi}_{nj}\left(\{{\bf W}_{ni}\}_{i=1}^j,\Theta_{nk},\lambda_{nnk}\right) \succeq {\bf 0},\forall m,n \in \mathcal{M}, j,k\in\mathcal{C}_n, j>k, \label{eq:admm1}\\
	&{\bf \Psi}_{nk}\left(\{{\bf W}_{ni}\}_{i=1}^k,\Theta_{nk},\lambda_{nnk}\right) \succeq {\bf 0},\forall m,n \in \mathcal{M}, k\in\mathcal{C}_n, \label{eq:admm2}\\
	&{\bf \Omega}_{mnk}\left(\{{\bf W}_{ni}\}_{i=1}^{N_n},\theta_{mnk},\lambda_{mnk}\right) \succeq {\bf 0},\forall m,n \in \mathcal{M}, k\in\mathcal{C}_n, \label{eq:admm3}\\	
	& \sum_{k=1}^{N_n} \tr({\bf W}_{nk}) \leq P_{\max}, \forall n\in\mathcal{M}, \\
	& p_n = \sum\nolimits_{k=1}^{N_n} {\bf W}_{nk}, \forall n \in \mathcal{M},\label{eq:admm4}\\
	& \Theta_{nk} = \sum\nolimits_{m=1,m\neq n}^{N_m} \theta_{mnk}, \forall n \in \mathcal{M}, k\in\mathcal{C}_n, \label{eq:admm5}
	\end{align}
\end{subequations}

To apply ADMM, one important step is to decompose the constraint set into several independent subsets. Fortunately, we observe that, with problem unchanged, the subindices $m$ and $n$ in \eqref{eq:admm3} can be interchanged. Hence, the constraints from \eqref{eq:admm1} to \eqref{eq:admm4} can be decomposed into the following $M$ independent convex sets
\begin{align} \label{eq:admm_set}
\mathcal{S}_n = \Big\{&\{{\bf W}_{nk}\},\{p_n\},\{\theta_{nmk}\},\{\Theta_{nmk}\},\{\lambda_{nmk}\}\big|\nonumber\\
&{\bf \Phi}_{nj}\left(\{{\bf W}_{ni}\}_{i=1}^j,\Theta_{nk},\lambda_{nnk}\right) \succeq {\bf 0}, \forall j,k\in\mathcal{C}_n, j>k, \nonumber\\
&{\bf \Psi}_{nk}\left(\{{\bf W}_{ni}\}_{i=1}^k,\Theta_{nk},\lambda_{nnk}\right) \succeq {\bf 0}, \forall k\in\mathcal{C}_n, \nonumber\\
&{\bf \Omega}_{nmk}\left(\{{\bf W}_{ni}\}_{i=1}^{N_n},\theta_{nmk},\lambda_{nmk}\right) \succeq {\bf 0}, \forall m\neq n, k\in\mathcal{C}_n, \nonumber\\
& \sum\nolimits_{k=1}^{N_n} \tr({\bf W}_{nk}) \leq P_{\max}, \nonumber\\
&{\bf W}_{nk} \succeq {\bf 0}, \lambda_{mnk} \geq 0,  \forall k\in\mathcal{C}_n, \nonumber\\
&p_n = \sum\nolimits_{k=1}^{N_n} {\bf W}_{nk}\Big\},~~ \forall n \in \mathcal{M},
\end{align}
Further, define the following variables:
\begin{subequations} \label{eq:ici_variables}
	\begin{align}
	{\bm{\theta}} &= \big[[\theta_{111},...,\theta_{11\tilde{K}}],...,[\theta_{MM1},...,\theta_{MM\tilde{K}}]\big] \in \Rset_+^{MM\tilde{K}}, \label{eq:ici}\\
	{\bm{\theta}}_n &= \big[[\Theta_{n1},...,\Theta_{n\tilde{K}}],[\theta_{n11},...,\theta_{n1\tilde{K}}],...,[\theta_{nM1},...,\theta_{nM\tilde{K}}]\big] \in \Rset_+^{M \tilde{K}}, n\in\mathcal{M}. \label{eq:ici_n} 
	\end{align}
\end{subequations}
where $\tilde{K} = \max\{N_m,\forall m\in \mathcal{M}\}$ and $\theta_{nmi} = 0$ for $N_n<i \leq \tilde{K}, \forall n,m\in \mathcal{M}$. \eqref{eq:ici} collects all the inter-cell interferences, while \eqref{eq:ici_n} contains $[\Theta_{n1},...,\Theta_{n\tilde{K}}]$ and $\{\theta_{nm\tilde{K}}\}_{m,\tilde{K}}$ (where $m\neq n$) that are relevant to UAV$_n$. Remind that $\Theta_{n\tilde{K}} = \sum_{m=1,m\neq n}^{N_m} \theta_{mn\tilde{K}}$, it can be verified that there exists a matrix ${\bf{E}}_n \in \{0,1\}^{M \tilde{K} \times MM\tilde{K}}$, such that ${\bm{\theta}}_n = {\bf{E}}_n {\bm{\theta}}, \forall n \in \mathcal{M}$. Then, based on \eqref{eq:admm_set} and \eqref{eq:ici_variables}, problem \eqref{eq:admm_p1} can be compactly rewritten as 

\begin{subequations}
	\begin{align}
	\min_{\{\{{\bf z}_n\}, {\bm \theta}\}} ~ &\sum_{n=1}^{M} p_n \\
	\st ~~& {\bf z}_n \triangleq \big( \{{\bf W}_{nk}\},\{p_n\},\{\theta_{nmk}\},\{\Theta_{nmk}\},\{\lambda_{nmk}\}\big) \in \mathcal{S}_n, \forall n \in \mathcal{M},\\
	&{\bm{\theta}}_n = {\bf{E}}_n {\bm{\theta}}, \forall n \in \mathcal{M}.
	\end{align}
\end{subequations}

Now, we are ready to apply ADMM. According to the rationale of ADMM, we solve the following augmented problem
\begin{subequations}\label{p:admm_augment}
	\begin{align}
	\min_{\{\{{\bf z}_n\}, {\bm \theta}\}} ~ &\sum\nolimits_{n=1}^{M} p_n + \frac{\rho}{2} ||{\bf{E}}_n {\bm{\theta}} - {\bm{\theta}}_n ||^2\\
	\st ~~& {\bf z}_n \in \mathcal{S}_n, \forall n \in \mathcal{M},\\
	&{\bm{\theta}}_n = {\bf{E}}_n {\bm{\theta}}, \forall n \in \mathcal{M}. \label{eq:admm_LagRG_ici}
	\end{align}
\end{subequations}
where $\rho > 0$ is a penalty parameter. As the first step of ADMM, we give the augmented Lagrangian
of \eqref{p:admm_augment} as follows:
\begin{align}\label{p:admm_augment_lagrg}
	L({\bf z}_n, {\bm \theta},{\bm{\nu}}_n) \triangleq \sum_{n =1}^{M} \left(p_n + \frac{\rho}{2} ||{\bf{E}}_n {\bm{\theta}} - {\bm{\theta}}_n ||^2 + {\bm{\nu}}_n ({\bf{E}}_n {\bm{\theta}} - {\bm{\theta}}_n)  \right)
\end{align}
where ${\bm \nu}_n$ is the dual variable associated with constraint \eqref{eq:admm_LagRG_ici}. Then, according to the principle of ADMM, we have the following primal updates:
	\begin{align}
	{\bf z}_n(q+1) &=  \arg \min_{{\bf z}_n \in \mathcal{S}_n} L({\bf z}_n, {\bm \theta}(q),{\bm{\nu}}_n(q)), \forall n \in \mathcal{M}, \label{eq:admm_z_update}\\
	{\bm \theta} (q+1) &= \arg \min_{{\bm \theta}} L({\bf z}_n(q+1), {\bm \theta},{\bm{\nu}}_n(q)),  \forall n \in \mathcal{M}, \label{eq:admm_theta_update}
	\end{align}
where $q$ is the iteration index. As it can be seen, problem \eqref{eq:admm_z_update} is convex, and thus, can be efficiently solved by {\tt{CVX}}. While for quadratic optimization problem \eqref{eq:admm_theta_update}, we have the following closed-form solution:
\begin{align} \label{p:admm_theta}
{\bm \theta}(q+1) = {\bf E}^{\dagger}\left( \tilde{\bm \theta}(q+1) - \tilde{\bm \nu}(q)/\rho\right)
\end{align}
where $\tilde{\bm \theta}(q+1)  = [{\bm \theta}_1^T(q+1),...,{\bm \theta}_{M}^T(q+1) ]^T$, $\tilde{\bm \nu}(q) = [{\bm \nu}_1^T(q),...,\tilde{\bm \nu}_M^T(q)]^T$ and ${\bf E} = [{\bf E}_1^T(q),...,$ $\tilde{\bf E}_M^T(q)]^T$.
Then, the dual variable ${\bm \nu}_n$ can be updated by 
\begin{align}\label{p:admm_nu}
{\bm \nu}_n(q+1) = {\bm \nu}_n(q) + \rho\big({\bf{E}}_n {\bm{\theta}} (q+1)- {\bm{\theta}}_n(q+1)\big)
\end{align}

The ADMM based algorithm proceed by iteratively updating ${\bf z}_n, \bm{\theta}$ and ${\bm \nu}_n$ until some convergence criteria is satisfied.
Finally, we outline the detailed steps of the ADMM based  decentralized robust beamforming design in Algorithm \ref{alg:admm}. Note that $\mathcal{S}_n, \forall n$, in \eqref{eq:admm_set} are bounded convex sets. Thus, according to \cite[Proposition 4.2]{boyd_admm}, Algorithm \ref{alg:admm} can guarantee to converge to the optimal solution of problem \eqref{eq:admm_p1}.

\begin{algorithm}[!tb] \smaller[1]
	\caption{ADMM based decentralized robust beamforming design for problem \eqref{p:sdp}}\label{alg:admm}
	\begin{algorithmic}[1]
		\STATE {{\bf Given} the initial values ${\bm \theta}(0)$, ${\bm{\nu}}_n(0)$ and the tolerance $\epsilon_0$.}\\
		\STATE {Set $q = 0$}
		\WHILE {$p_n(q+1) - p_n(q) > \epsilon_0$}
		\STATE {UAV $m$ solves the local problem \eqref{eq:admm_z_update} to update the primal variables ${\bf z}_n$.}
		\STATE {UAV $m$ reports ${\bm \theta}_n$ to the other UAVs.}
		\STATE {UAV $m$ updates the primal variables $\bm \theta$ by solving \eqref{p:admm_theta}.}\\
		\STATE {UAV $m$ updates the dual variables $\bm \nu$ by solving \eqref{p:admm_nu}.}\\
		\STATE {Set $q := q+1$.}\\
		\ENDWHILE
		\STATE {{\bf{Output :}} The beamfoming vectors $\{{\bf W}_{nk}\}$}
	\end{algorithmic}
\end{algorithm}

\section{Simulation Results} \label{sec:simulation}
In this section, numerical simulations are present to verify the performance of the proposed transmission schemes and algorithms.
Following the parameter setup in \cite{LTE-Sim}, the large-scale path loss is set to be $PL_{mnk} = 128.1 + 37.6\log_{10}(d_{mnk})$ with $d_{mnk}$ (in Km) denoting the distance between the UAV$_m$ and U$_{nk}$. The channels between a UAV and its neighbouring users are modelled by standard Rayleigh fading, while  the channels between a UAV and its home users are characterized by the Racian fading. Thus, we have 
\begin{align}
	{\bf h}_{nnk} &= \sqrt{\frac{K_r}{1+K_r}}{\bf h}_{nnk}^{\rm LoS} + \sqrt{\frac{1}{1+K_r}}{\bf h}_{nnk}^{\rm NLoS},\forall n \in \mathcal{M}, k\in\mathcal{C}_n,
\end{align}
where $K_r = 3$ denotes the Rician factor, ${\bf h}_{nnk}^{\rm LoS}$ follows the LoS deterministic component, and ${\bf h}_{nnk}^{\rm NLoS}$ is the standard Rayleigh fading component. The White noise power density is $-174$dBm and the bandwidth is $10$MHz. The UAV hovers at
a fixed altitude that is set as $h_n = 100{\rm m},\forall n \in \mathcal{M}$. The power budget of each UAV is set as $P_{\max} = 36$dBm. The users are deployed in a $500$m $\times$ $500$m square area following the PCP with $r_n = r = 50$m, $\forall n \in \mathcal{M}$. Without loss of generality, the numbers of users in each cluster are assumed to be equal, i.e., $N_n = N, \forall n \in \mathcal{M}$ and the QoS requirements of the users are the same, i.e., $\gamma_{mnk} = \gamma, \forall m,n\in \mathcal{M},k\in\mathcal{C}_n$. The spherical error model is used and we assume that the error bounds of all users are the same, i.e., ${\bf Q}_{mnk} = {\bf Q} = \frac{1}{\epsilon^2}{\bf I}_{N_t}, \forall m,n\in\mathcal{M},\forall k\in\mathcal{C}_n$.

For comparison, 
we introduce some other user clustering approaches and transmission schemes, namely, the exhaustive search based user clustering approach, Swap-matching based user clustering approach, OMA transmission scheme, and non-robust NOMA transmission scheme, which 
are described as follow:
\begin{itemize}
	\item {\bf Exhaustive Search Based User Clustering Approach:} In this approach, all the possible combinations of user cluster are considered, the optimal user clustering is the one yielding the smallest transmit power.
	\item {\bf Swap-Matching Based User Clustering Approach:}  In this approach, the users are clustered using the swap-matching based algorithm \cite{xu-tcom-harq,di-user-pairing}. We will give a sketch of the swap-matching based algorithm here, and the interested reader can refer to \cite{xu-tcom-harq,di-user-pairing} for the details. The swap-matching based user clustering algorithm consists of two phases, i.e., the initial matching phase and the swap matching phase. In the initial matching phase, $M$ users are first selected as proposers. Then, each proposer will select $N-1$ users that can provide better performance and meanwhile are not selected by other proposers to form a user cluster. By doing this, the initial clustering outcome is established. In the swap matching phase, each proposer will perform swapping operations to exchange the users in the same cluster with the other proposers. If the swapping operation can decrease the consumed power of the relevant two clusters and doesn't hurt the performance of other clusters, the swapping operation will be approved. The swapping operation will continue until the clustering is stable.
	\item {\bf OMA Transmission Scheme:} In this scheme, the time division multiple access (TDMA) is used. The transmission duration is evenly divided for the users in the same cluster. Compared to the proposed NOMA transmission scheme, the interferences solely come from the inter-clusters and the constraints on SIC are removed. Thus, the problem of the OMA transmission scheme is actually a simplified version of the NOMA transmission scheme, and the proposed algorithm can be adopted to solve the problem of the OMA transmission scheme.
	\item {\bf Non-robust NOMA transmission scheme:} In this scheme, perfect CSI of each user is assumed to be available at the UAVs. As studied in subsection \ref{subsec:non-robust}, the relevant problem can be formulated as an SOCP and thus can be solved optimally.
\end{itemize}

\begin{figure}[!tp]\centering
	\includegraphics[width=0.66\linewidth]{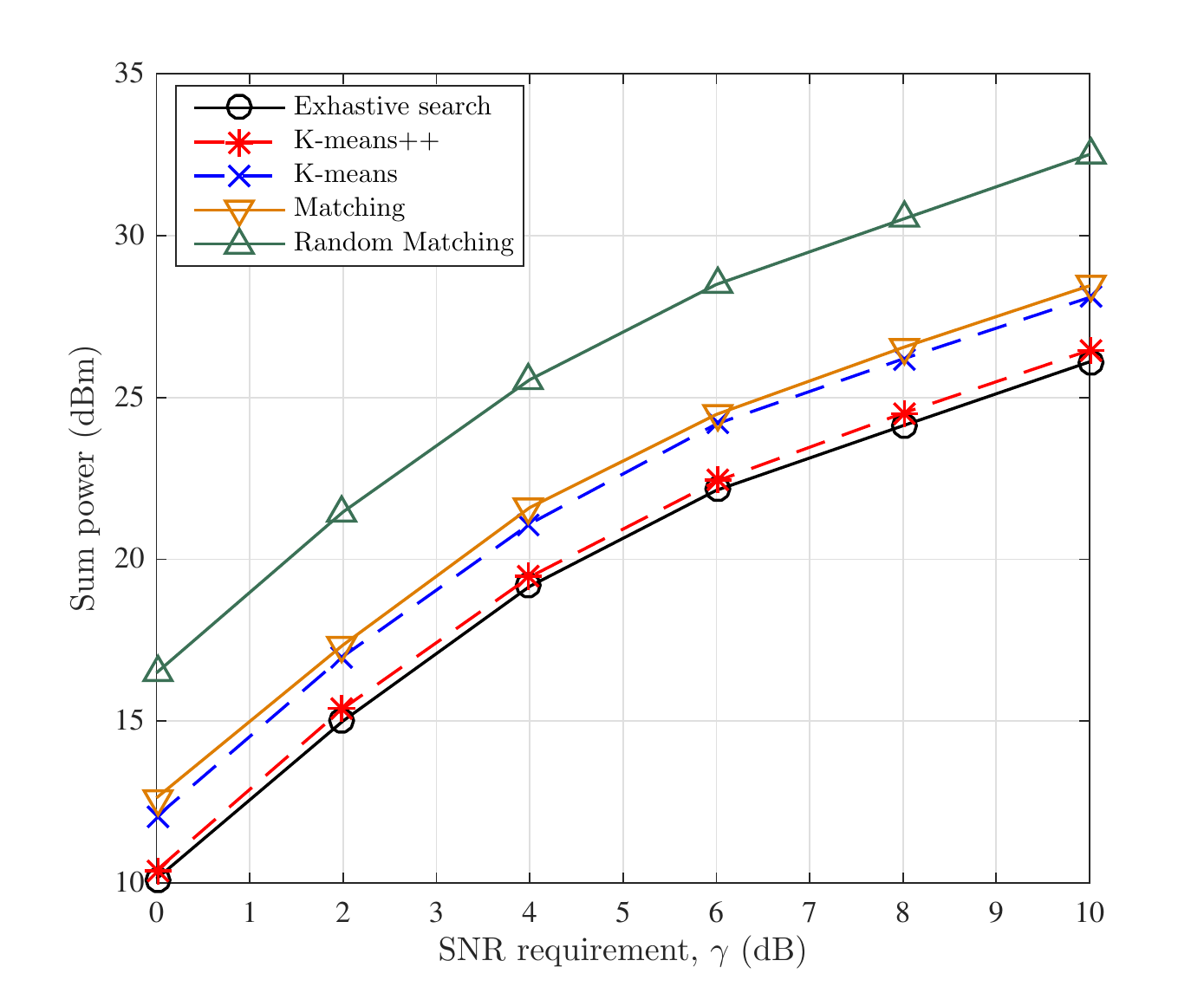}
	\caption{The performance comparison of the centralized and decentralized algorithm with $M=4$, $N=3$, $N_t = 4$, $\epsilon = 0.05$ and $\gamma = 3$dB.}  \label{fig:power_clustering}
\end{figure}

\begin{figure}[!tp]\centering
	\includegraphics[width=0.66\linewidth]{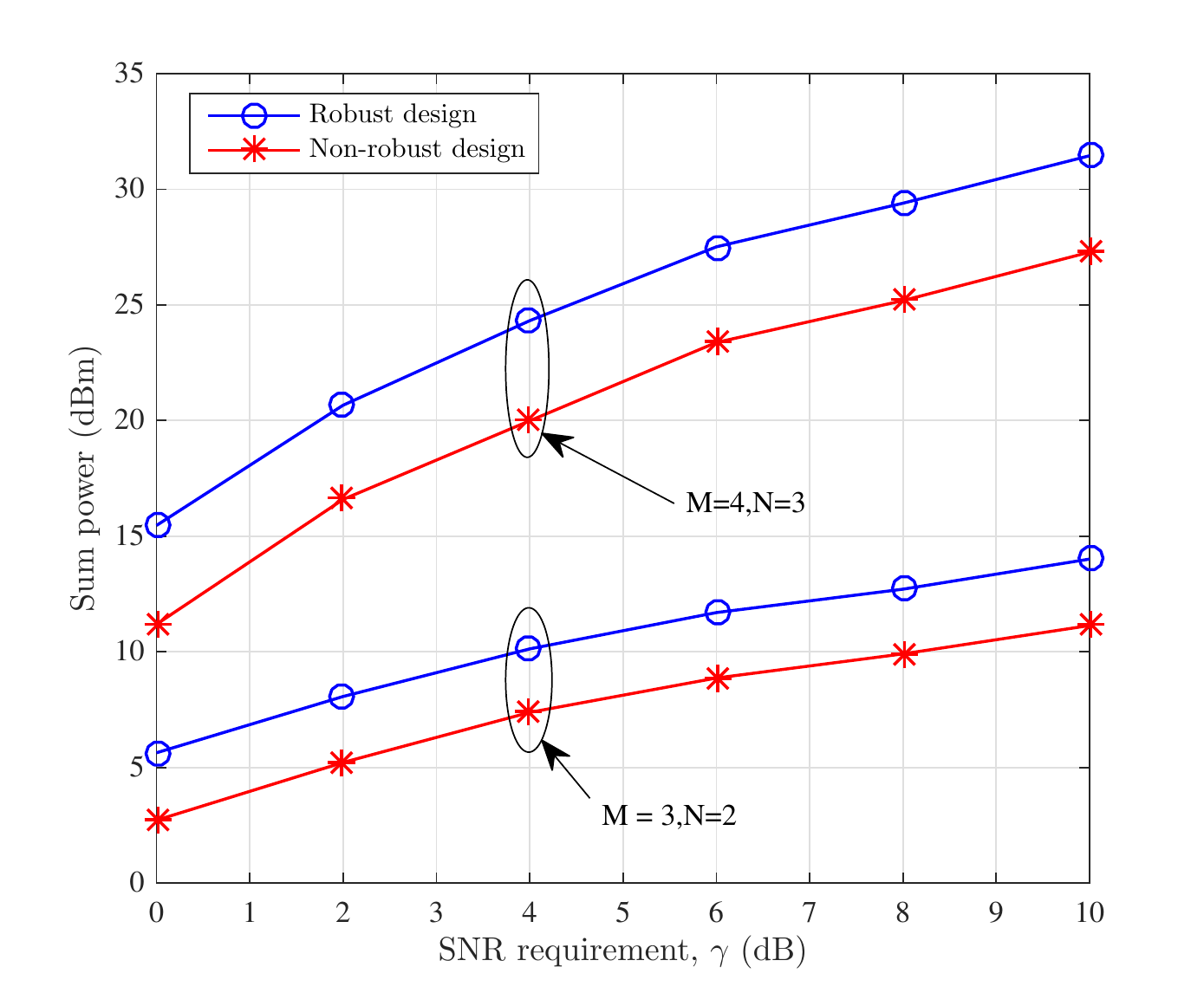}
	\caption{The performance comparison of robust and non-robust designs versus different $\gamma$ with $N_t = 4$, and $\epsilon = 0.05$.}  \label{fig:power_robust_gamma}
\end{figure}

\subsection{Performance Evaluation of the Proposed K-means++ Based User Clustering Algorithm}
We first evaluate the performance of the proposed k-means++ based user clustering approach. The exhaustive search based optimal user clustering algorithm is used as the performance benchmark. As it can be seen that the k-mean++ based approach can achieve the near optimal performance under the system settings. Note that there is still a gap between the performances of k-mean++ based approach and the optimal one, this is due to reason that the k-means++ cannot guarantee to find the appropriate initial cluster centroids always.
We can also observe that the k-means++ based approach outperforms the standard k-means based approach due to its careful initial cluster centroids selection. Meanwhile, the swap-matching based approach can perform closely to the k-means based approach. While the random matching based approach, in which the users are randomly selected to form a cluster, yields the worst performance. 

\subsection{Performance Comparison of the Robust and Non-Robust Designs}
\begin{figure}[!tp]\centering
	\includegraphics[width=0.66\linewidth]{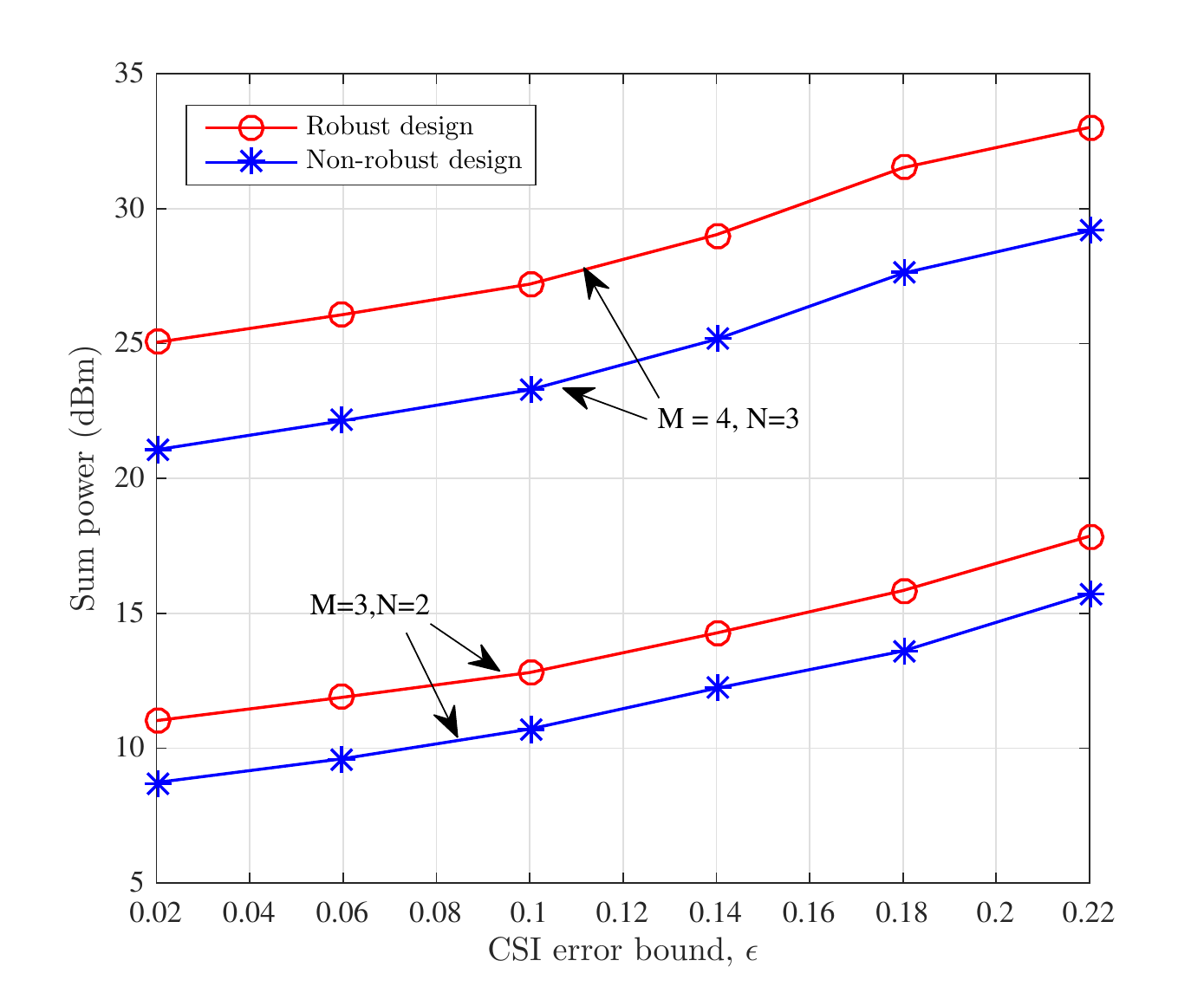}
	\caption{The performance comparison of robust and non-robust designs versus different $\epsilon$ with $N_t = 4$,  and $\gamma = 3$dB.}  \label{fig:power_robust_csi}
\end{figure}

\begin{figure}[!tp]\centering
	\includegraphics[width=0.66\linewidth]{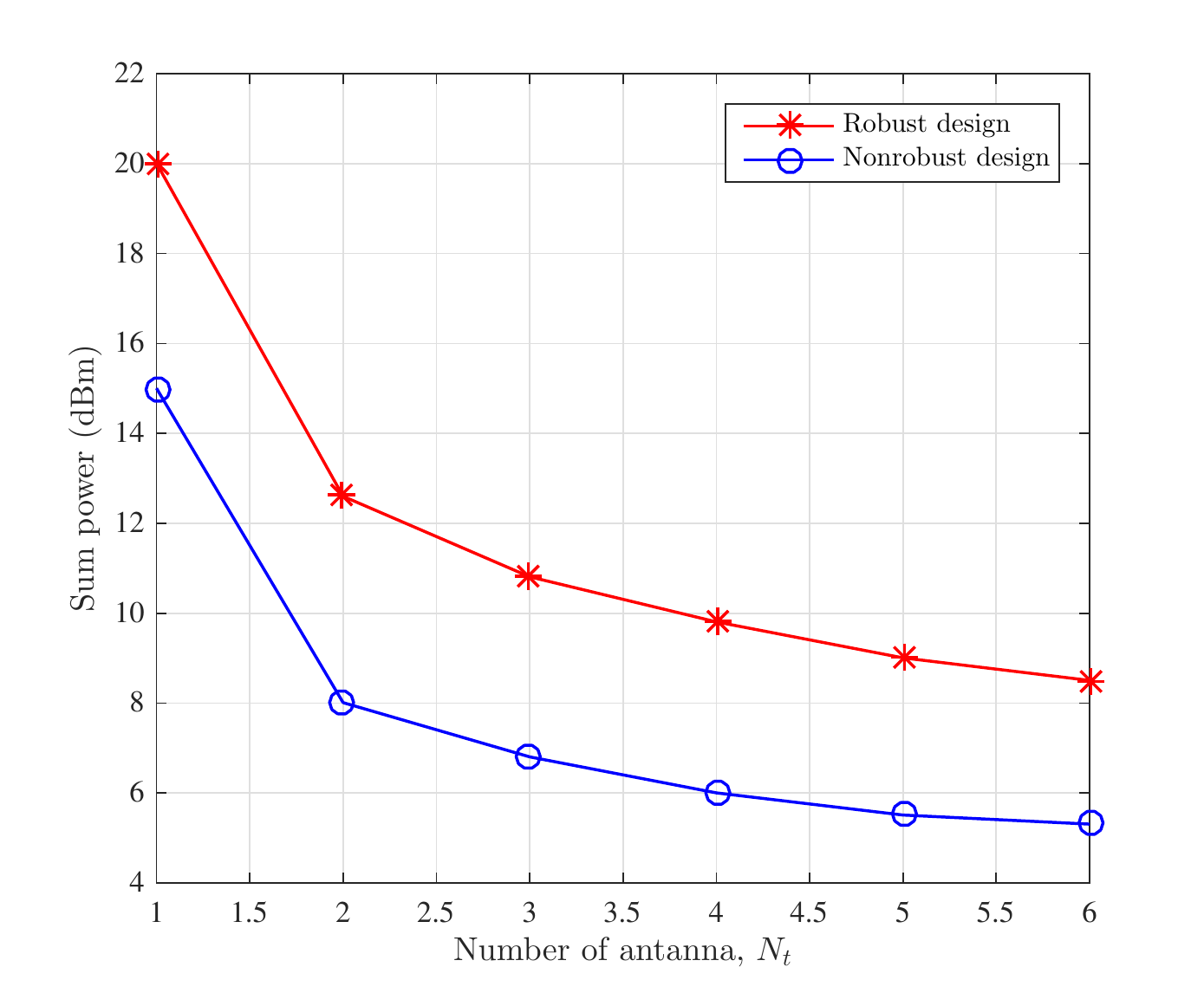}
	\caption{The performance comparison of robust and non-robust designs versus different number of antennas with $M=3$, $N=2$, $\epsilon = 0.05$, and $\gamma = 3$dB.}  \label{fig:power_nt}
\end{figure}

In this subsection, we evaluate the performance of the robust design, with the non-robust design serving as the benchmark. Notice that, in the non-robust design, the CSI is assumed to be perfectly known at the UAVs. As it is shown that the associated problem can be formulated as an SOCP and thus can be optimally solved by standard convex solver. In Fig. \ref{fig:power_robust_gamma}, we compare the performance of robust and non-robust design versus different QoS requirements of users. It can be observed that as a price for worst-case performance guarantee, the robust designs require higher transmission power than the non-robust design. 
From Fig. \ref{fig:power_robust_csi}, we can also find that, with different CSI error bounds, the non-robust design would underestimate the required power for reliable transmissions. Meanwhile, the sum power increases with the increase of the CSI error bound. The reason is that, for a larger CSI error, it requires more transmission power to guarantee the QoS requirements of users.
In Fig. \ref{fig:power_nt}, the sum power consumptions of robust and nonrobust designs with different number of antennas are compared. It can be observed that, by benefiting the diversity gain of the multi-antenna technique, the consumed power decreases with the increases of the number of antennas.

\subsection{Performance Comparison of the NOMA and OMA Transmission Schemes}
\begin{figure}[!tp]\centering
	\includegraphics[width=0.66\linewidth]{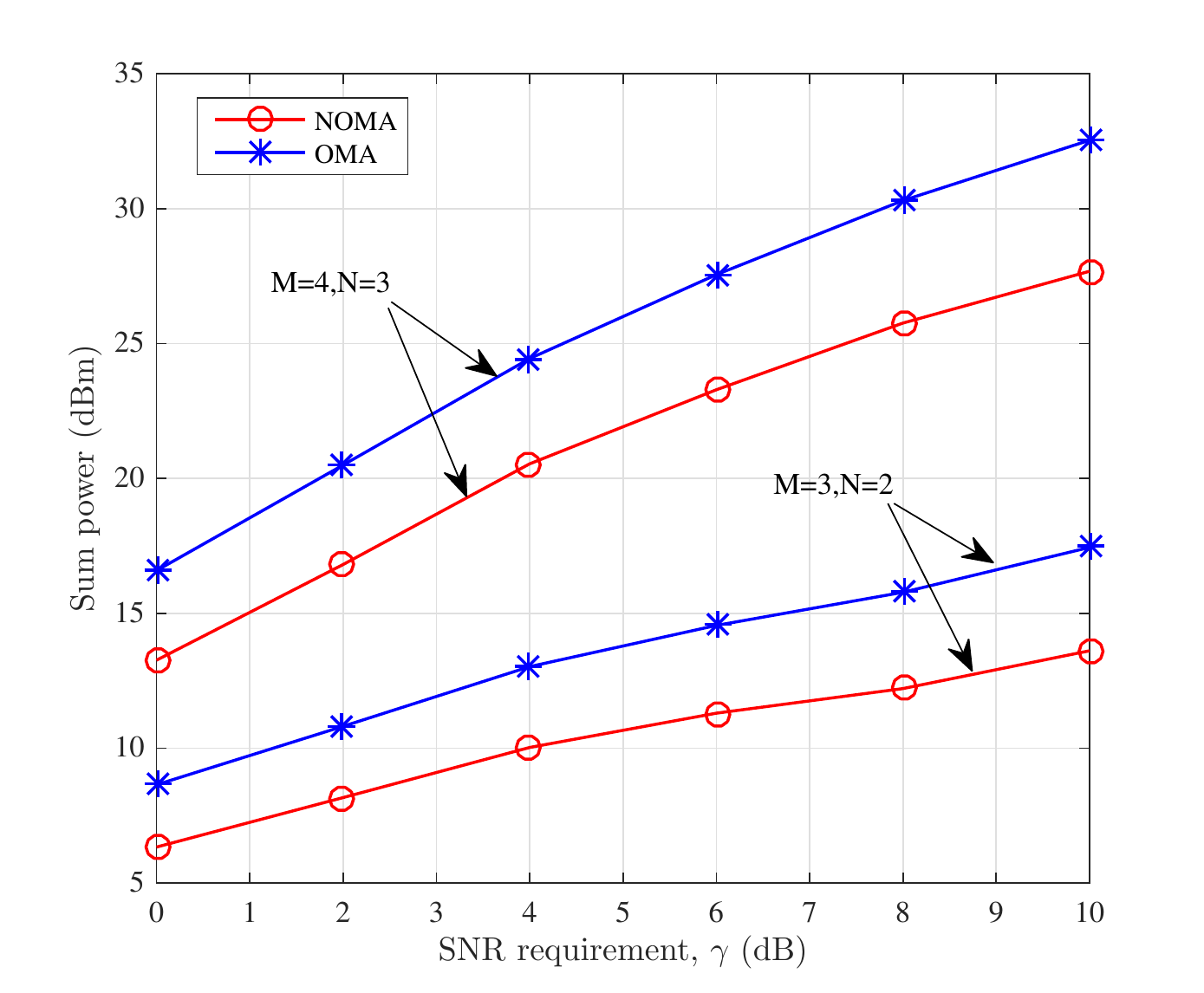}
	\caption{The performance comparison of NOMA and OMA transmission schemes versus different $\gamma$ with $N_t = 4$, and $\epsilon = 0.05$.}  \label{fig:power_ma_gamma}
\end{figure}

In Fig. \ref{fig:power_ma_gamma}, the performances of the proposed robust NOMA transmission scheme and the robust TDMA transmission scheme are compared with different SNR requirements. As it can be observed that, due to the superior spectral efficiency, the proposed NOMA transmission scheme can out perform the TDMA scheme and the performance gain of the proposed NOMA scheme increases with the increase of the SNR requirement.

\subsection{Performance Comparison of the Centralized and Decentralized Algorithms}
\begin{figure}[!tp]\centering
	\includegraphics[width=0.66\linewidth]{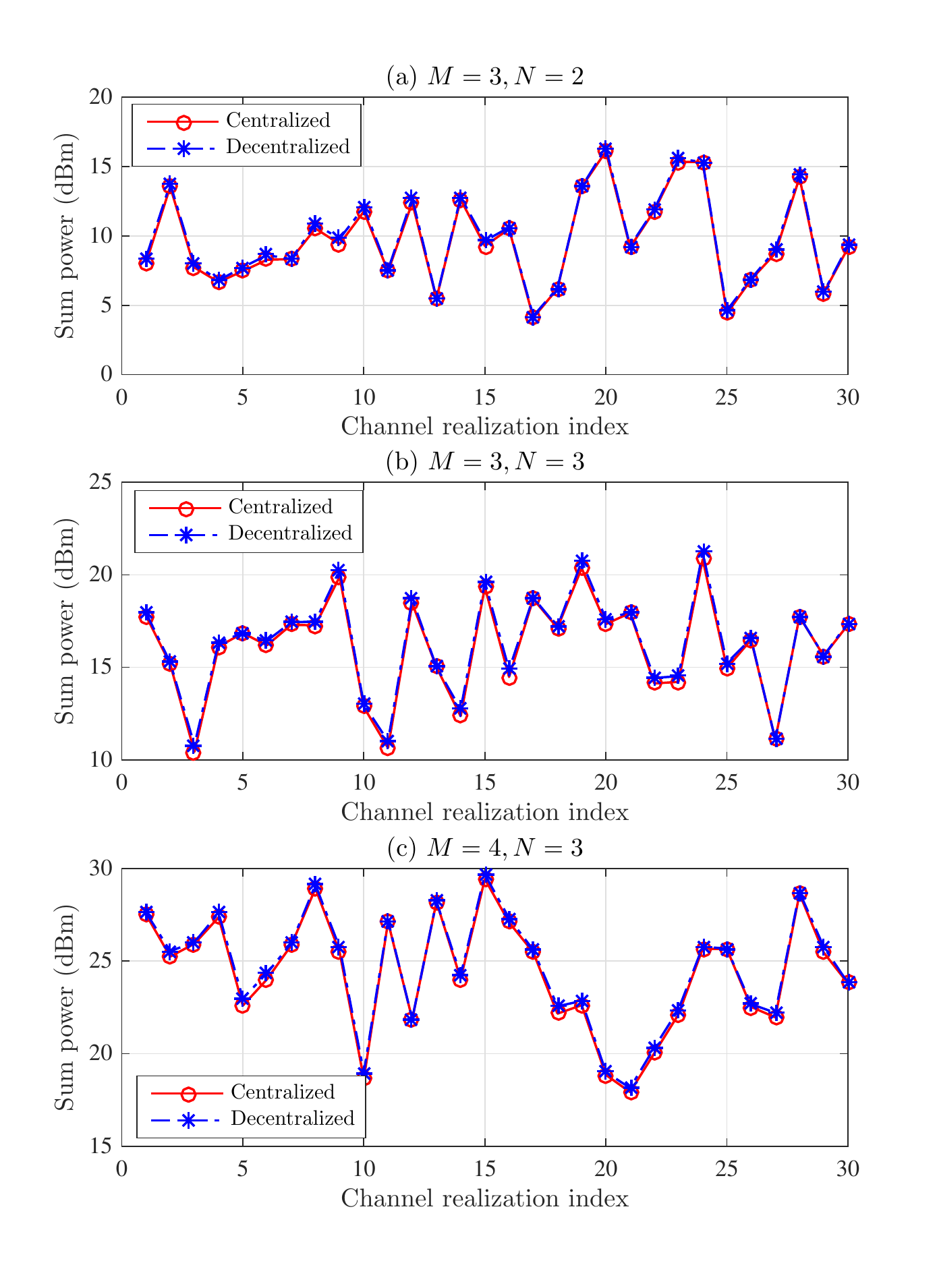}
	\caption{The performance comparison of the centralized and decentralized algorithm with $N_t = 4$, $\epsilon = 0.05$ and $\gamma = 3$dB.}  \label{fig:centr-decentr}
\end{figure}

To compare the performance of the centralized algorithm, by solving problem \eqref{p:sdp} directly with {\tt CVX}, and the decentralized algorithm (i.e., Algorithm \ref{alg:admm}), we evaluate the sum power consumption with different number of user clusters and users in each user cluster over 30 randomly generated channel realizations. As it can be seen from Fig. \ref{fig:centr-decentr} that, in most cases, the ADMM based decentralized algorithm can perform closely to the centralized algorithm.

\section{Conclusions} \label{sec:conclusion}
In this work, we have investigated the problem of joint user clustering and robust beamforming design
in a downlink NOMA network with multi-users and multi-UAVs under imperfect CSI assumptions. The formulated sum power minimization problem was shown to be challenging to solve due to the integer variables of user clustering, coupling effects of beamformers, and infinitely many constraints bringing by the imperfect CSI. 
For computational complexity reduction, the original problem was decoupled into user clustering subproblem and robust beamforming subproblem.
By utilizing the users' position information, we proposed a k-means++ based unsupervised clustering algorithm to solve the user clustering problem. Then, the robust beamforming design problem was considered. Firstly, the problem with perfect CSI assumptions was globally solved by transforming the relevant problem into an SOCP. Then, we proposed an SDR based suboptimal algorithm to solve the robust beamforming design problem in the general case with imperfect CSI assumption, and a sufficient condition under which the SDR based approach could guarantee to produce an optimal solution was presented. Finally, an ADMM based decentralized algorithm was developed to allow the UAVs to determine the beamforming design by using local CSI of its home users. Simulation results have provided some interesting results. For example, the k-means++ based algorithm can outperforms the standard k-means based algorithm and traditional swap matching based algorithm; The robust design would require more transmission power compared to the non-robust design; and the proposed NOMA transmission
scheme can greatly outperform the TDMA transmission scheme, especially for the case in which the users have high QoS requirements.

\begin{appendices}
	\section{Proof of Lemma \ref{lem:s-lemma}} \label{apdx:s-lemma}
	First, note that the channels, and thus the CSI errors, in the left- and right-hand-sides of \eqref{eq:sdr_sic} and \eqref{eq:sdr_qos} are independent. Hence, according to \eqref{eq:sdr_sic} and \eqref{eq:sdr_qos}, the SIC and QoS constraints for U$_{nk}$ can be equivalently rewritten as
	\begin{subequations} 
		\begin{align}
		& \min_{\forall {\bf e}^H_{nnk} {\bf Q}_{nnk} {\bf e}_{nnk} \leq 1} \left\{{\bf h}^H_{nnk} \left(\frac{1}{\gamma_{nj}}{\bf W}_{nj} - \sum_{i=1}^{j-1} {\bf W}_{ni}\right){\bf h}_{nnk}\right\} \geq \notag\\
		&\quad\quad\quad\quad\quad\quad\quad\quad\max_{\forall {\bf e}^H_{mnk} {\bf Q}_{mnk} {\bf e}_{mnk} \leq 1,} \left\{\sum_{m=1,m\neq n}^M {\bf h}^H_{mnk} \left(\sum_{i=1}^{N_m} {\bf W}_{mi}\right) {\bf h}_{mnk} \right\} + \sigma_{nk}^2, \label{eq:ap_sdr_sic}\\
		& \min_{\forall {\bf e}^H_{nnk} {\bf Q}_{nnk} {\bf e}_{nnk} \leq 1} \left\{{\bf h}^H_{nnk} \left(\frac{1}{\gamma_{nk}}{\bf W}_{nk} - \sum_{i=1}^{k-1} {\bf W}_{ni}\right){\bf h}_{nnk}\right\} \geq \notag\\
		&\quad\quad\quad\quad\quad\quad\quad\quad\max_{\forall {\bf e}^H_{mnk} {\bf Q}_{mnk} {\bf e}_{mnk} \leq 1,} \left\{\sum_{m=1,m\neq n}^M {\bf h}^H_{mnk} \left(\sum_{i=1}^{N_m} {\bf W}_{mi}\right) {\bf h}_{mnk} \right\}  + \sigma_{nk}^2,\label{eq:ap_sdr_qos}
		\end{align}
	\end{subequations}
 By introducing slack variables
 \begin{align}
 \theta_{mnk} = \max_{\forall {\bf e}^H_{mnk} {\bf Q}_{mnk} {\bf e}_{mnk} \leq 1,} {\bf h}^H_{mnk} \left(\sum_{i=1}^{N_m} {\bf W}_{mi}\right) {\bf h}_{mnk},
 \end{align}
 and based on \eqref{eq:ap_sdr_sic} and \eqref{eq:ap_sdr_qos}, the worst-case SINR constraints for
 in \eqref{eq:sdr_sic} and \eqref{eq:sdr_qos} can be decoupled into the following $M+1$ worst-case constraints
	\begin{subequations} \label{eq:worse_case_csi}
	\begin{align}
	& \min_{\forall {\bf e}^H_{nnk} {\bf Q}_{nnk} {\bf e}_{nnk} \leq 1} \left\{{\bf h}^H_{nnk} \left(\frac{1}{\gamma_{nj}}{\bf W}_{nj} - \sum_{i=1}^{j-1} {\bf W}_{ni}\right){\bf h}_{nnk}\right\} \geq \sum_{m=1,m\neq n}^M \theta_{mnk} + \sigma_{nk}^2,\\
	& \min_{\forall {\bf e}^H_{nnk} {\bf Q}_{nnk} {\bf e}_{nnk} \leq 1} \left\{{\bf h}^H_{nnk} \left(\frac{1}{\gamma_{nk}}{\bf W}_{nk} - \sum_{i=1}^{k-1} {\bf W}_{ni}\right){\bf h}_{nnk}\right\} \geq \sum_{m=1,m\neq n}^M \theta_{mnk} + \sigma_{nk}^2,\\
	&  \theta_{mnk} \geq {\bf h}^H_{mnk} \left(\sum_{i=1}^{N_m} {\bf W}_{mi}\right) {\bf h}_{mnk}, \forall {\bf e}^H_{mnk} {\bf Q}_{mnk} {\bf e}_{mnk} \leq 1, \forall m \in \mathcal{M} \backslash \{n\} 
	\end{align}
\end{subequations}
Note that each of the constraints in \eqref{eq:worse_case_csi} contains only one CSI error. Thus, the \textit{S-Lemma} \cite{Boyd-2009} can be applied to reformulate \eqref{eq:worse_case_csi} into the a finite number of constraints that are given in \eqref{eq:lemma1}.
This completes the proof. 	\hfill $\blacksquare$

\section{Proof of Lemma \ref{lem:rank-one}} \label{app:rank}
In this appendix, we proof that $\Rank\left({\bf W}_{nk}\right) = 1, \forall n,k$. As perfect intra-cell CSI errors are available at the UAVs, problem \eqref{p:sdr_robust} degrades to
\begin{subequations}\label{p:sdr_robust2}
	\begin{align}
	\min_{\{{\bf W}_{nk}\}_{\forall n,k}}~ &\sum_{n=1}^{M} \sum_{k=1}^{N_n} \tr({\bf W}_{nk})\\
	\st~~~&\hat{\bf h}^H_{nnk} \left(\frac{1}{\gamma_{nj}}{\bf W}_{nj} - \sum_{i=1}^{j-1} {\bf W}_{nk}\right)\hat{\bf h}_{nnk} \geq \notag\\
	&\quad\max_{\forall {\bf e}^H_{mnk} {\bf Q}_{mnk} {\bf e}_{mnk} \leq 1,} \left\{\sum_{m=1,m\neq n}^M \hat{\bf h}^H_{mnk} \left(\sum_{i=1}^{N_m} {\bf W}_{mi}\right) \hat{\bf h}_{mnk}\right\}  + \sigma_{nk}^2, \forall , k\in \mathcal{C}_n,m,n\in\mathcal{M}, \label{eq:sdr_sic2}\\
	& \hat{\bf h}^H_{nnk} \left(\frac{1}{\gamma_{nk}}{\bf W}_{nk} - \sum_{i=1}^{k-1} {\bf W}_{ni}\right)\hat{\bf h}_{nnk} \geq \notag\\
	&\quad\max_{\forall {\bf e}^H_{mnk} {\bf Q}_{mnk} {\bf e}_{mnk} \leq 1,} \left\{\sum_{m=1,m\neq n}^M \hat{\bf h}^H_{mnk} \left(\sum_{i=1}^{N_m} {\bf W}_{mi}\right) \hat{\bf h}_{mnk}\right\}  + \sigma_{nk}^2, \forall , k\in \mathcal{C}_n,m,n\in\mathcal{M}, \label{eq:sdr_qos2}\\	
	& \sum_{k=1}^{N_n} \tr({\bf W}_{nk}) \leq P_{\max}, \forall n\in\mathcal{M}, 
	\end{align}
\end{subequations}
Again, by applying {\textit{S-Lemma}} to the right-hand-sides of \eqref{eq:sdr_sic2} and \eqref{eq:sdr_qos2}, problem \eqref{p:sdr_robust2} can be equivalently reformulated as:
\begin{subequations}\label{p:sdr_robust3}
	\begin{align}
	&\!\!\!\!\!\!\!\!\!\min_{\{{\bf W}_{nk}\},v_{mnk},\lambda_{mnk}}~ \sum_{n=1}^{M} \sum_{k=1}^{N_n} \tr({\bf W}_{nk})\\
	\st~&\hat{\bf h}^H_{nnk} \left(\frac{1}{\gamma_{nj}}{\bf W}_{nj} - \sum_{i=1}^{j-1} {\bf W}_{nk}\right)\hat{\bf h}_{nnk} \geq \sum_{m=1,m\neq n} v_{mnk} + \sigma_{nk}^2, \forall , k\in \mathcal{C}_n,m,n\in\mathcal{M}, \label{eq:sdr_sic3}\\
	& \hat{\bf h}^H_{nnk} \left(\frac{1}{\gamma_{nk}}{\bf W}_{nk} - \sum_{i=1}^{k-1} {\bf W}_{ni}\right)\hat{\bf h}_{nnk} \geq\sum_{m=1,m\neq n} v_{mnk} + \sigma_{nk}^2, \forall , k\in \mathcal{C}_n,m,n\in\mathcal{M},\label{eq:sdr_qos3}\\
	&\left[
	\begin{matrix}
	-\sum\limits_{i=1}^{N_m} {\bf W}_{mi} \!+\!\lambda_{mnk}{\bf Q}_{mnk} & -\sum\limits_{i=1}^{N_m} {\bf W}_{mi}\hat{\bf h}_{mnk}\\
	-\hat{\bf h}_{mnk}^H \sum\limits_{i=1}^{N_m} {\bf W}_{mi} & -\hat{\bf h}_{mnk}^H \sum\limits_{i=1}^{N_m} {\bf W}_{mi} \hat{\bf h}_{mnk} \!+\! v_{mnk} \!+\! \lambda_{mnk}
	\end{matrix}
	\right] \succeq {\bf 0},\forall m,n \in \mathcal{M}, k\in\mathcal{C}_n, \label{eq:sdr_app1}\\
	& \sum_{k=1}^{N_n} \tr({\bf W}_{nk}) \leq P_{\max}, \forall n\in\mathcal{M}, \\
	& {\bf W}_{nk} \succeq {\bf 0}, \forall n \in \mathcal{M}, k\in\mathcal{C}_n. \label{eq:sdr_app2}
	\end{align}
\end{subequations}

We prove Lemma \ref{lem:rank-one} by using the KKT conditions of \eqref{p:sdr_robust3}. In particular, let $\{\delta_{nj}^{\star}\}$, $\{\epsilon_{nj}^{\star}\}$, ${\bf Y}_{mi}^{\star}$ and ${\bf Z}_{nk}^{\star}$ denote the optimal dual variables associated with \eqref{eq:sdr_sic3}, \eqref{eq:sdr_qos3}, \eqref{eq:sdr_app1} and \eqref{eq:sdr_app2}, respectively. The KKT conditions related to ${\bf W}_{nk}^{\star}$ are as follows:
\begin{subequations}
	\begin{align}
	&{\bf Z}_{nk}^{\star} {\bf W}_{nk}^{\star} = {\bf 0},\\
	&{\bf Z}_{nk}^{\star} = {\bf I}_{N_t} - \left(\frac{\delta_{nk}^{\star}}{\gamma_{nj}} + \frac{\epsilon_{nk}^{\star}}{\gamma_{nk}}\right) {\hat{\bf h}}_{nnk}{\hat{\bf h}^H}_{nnk},\\
	& {\bf Z}_{nk}^{\star} \succeq {\bf 0}, {\bf W}_{nk}^{\star} \succeq {\bf 0}
	\end{align}
\end{subequations}
First, note that, from \eqref{eq:sdr_qos3}, we can conclude that ${\bf W}_{nk}^{\star} \succ {\bf 0}$. Otherwise, we have 
$$-\hat{\bf h}^H_{nnk}\sum_{i=1}^{k-1} {\bf W}_{ni}\hat{\bf h}_{nnk}  \geq\sum_{m=1,m\neq n} v_{mnk} + \sigma_{nk}^2,$$
which violates the fact that $v_{mnk}\geq 0,~ \sigma_{nk}^2 >0$ and ${\bf W}_{ni}$ are positive semidefinite. Then, also note that 
\begin{align}\label{eq:rank_z}
0 = \Rank\left({\bf Z}_{nk}^{\star} {\bf W}_{nk}^{\star}\right) \geq \Rank\left({\bf Z}_{nk}^{\star}\right) + \Rank\left({\bf W}_{nk}^{\star}\right) - N_t.
\end{align}
So, we have $\Rank\left({\bf W}_{nk}^{\star}\right) \leq  N_t - \Rank\left({\bf Z}_{nk}^{\star}\right)$. Therefore, to prove that ${\bf W}_{nk}$ is rank-one, it suffices to prove that 
\begin{align}
\Rank\left({\bf Z}_{nk}^{\star}\right)  &= \Rank\left({\bf I}_{N_t} - \left(\frac{\delta_{nk}^{\star}}{\gamma_{nj}} + \frac{\epsilon_{nk}^{\star}}{\gamma_{nk}}\right) {\hat{\bf h}}_{nnk}{\hat{\bf h}}_{nnk}^H\right) \nonumber\\
& = N_t -1. 
\end{align}
Notice that ${\hat{\bf h}}_{nnk}$  is a non-zero vector, thus $\Rank\left({\hat{\bf h}}_{nnk}{\hat{\bf h}}_{nnk}^H\right) = 1$. So, we have $\Rank\left({\bf Z}_{nk}^{\star}\right) \geq N_t-1$. Finally,  based on \eqref{eq:rank_z} and ${\bf W}_{nk}^{\star} \succ {\bf 0}$, we can conclude that $\Rank\left({\bf Z}_{nk}^{\star}\right) = N_t-1$ and $\Rank\left({\bf W}_{nk}^{\star}\right) = 1$. This completes the proof. 	\hfill $\blacksquare$

\end{appendices}




\end{document}